\documentclass[twocolumn,preprintnumbers,amsmath,amssymb,superscriptaddress]{revtex4}

\usepackage{setspace}

\usepackage{graphicx}
\usepackage{graphicx,epsf} 
\usepackage{dcolumn}
\usepackage{bm}
\usepackage{latexsym}
\newcommand{\xsize}{\epsfxsize=9.0cm}

\begin{document}

\title{A Complexity O(1) Priority Queue for Event Driven Molecular Dynamics 
Simulations }

\author{Gerald~Paul}
\affiliation{Center for Polymer Studies and Dept.\ of Physics, Boston
  University, Boston, MA 02215, USA} 
\email{gerryp@bu.edu}

\begin{abstract}

We propose and implement a priority queue suitable for use in event
driven molecular dynamics simulations.  All operations on the queue take
on average O(1) time per collision.  In comparison, previously studied
queues for event driven molecular dynamics simulations require O(log
$N$) time per collision for systems of $N$ particles.

\end{abstract}  

\maketitle

\section{Introduction}

Molecular dynamics simulations are a powerful tool for determining the
behavior of multiparticle systems and are used in a wide range of
applications \cite{Allen87,Becker,Daggett,Frenkel,Rapaport,Sadus,Urbanc}.
  
There are two basic approaches to these simulations:

(i) Time driven simulations \cite{Allen87} in which equations of motion of all
particles are solved for a series of small time slices.  The positions and 
velocities of the particles are determined at
the end of each time slice and used as input to the calculation for the 
next time slice.  

(ii) Event driven simulations
\cite{Alder,Allen89,Erpenbeck,Krantz,Donev1} which are applicable to
systems of hard spheres or more generally to systems with interparticle
potentials which are piecewise constant.  The approach with event driven
simulations is to determine when the next collision between two
particles occurs, determine the positions and velocities of these
particles after the collision and then repeat this process.  A collision
is defined as the event in which the hard spheres collide or more
generally when two particles reach a discontinuity in their
interparticle potential.

We focus here on event driven simulations which, where applicable,
provide exact results and typically run faster than time driven
simulations.  Determination of the next event is usually composed of two
steps \cite{Krantz}:

(i) determination of the collision event with the shortest time for
each particle.  By dividing the system into cells and/or maintaining
lists of particles within a certain distance of a given
particle (neighbor lists), the time taken for calculation of the first
collision event for a given particle can be made independent of N, the
 total number of particles in the system \cite{Erpenbeck,Donev1}.

(ii) determination of the collision event with the shortest time among
 all the particles, given the events with the shortest time for each
 particle obtained in (i). Approaches have been proposed and implemented
 which allow this determination in O($\log N$) time.

The subject of this paper is an approach to determining the next
collision event among all particles.  This has been a heavily researched
subject \cite{Lubachevsky,Marin93,Marin,Rapaport80,Shida}.  The
requirements for a queue to allow this determination is as follows. The
queue must support:

(i) addition of an event to the queue;

(ii) identification and deletion from the queue of the event with the
shortest collision time;

 (iii) deletion of a given event from the queue (e.g. when a collision
(p,q) occurs we may want to remove the event (q,p) from the queue.)
These requirements define abstractly the concept of a {\it priority
queue}.

Implementations of priority queues for molecular dynamics
simulations have for the most part been based on various types of binary
trees. They all share the property that determining the event in the
queue with the smallest value requires O($\log N$) time \cite{Marin}.

The early work on priority queues is reviewed in
Ref.~\cite{Jones}.  The earliest implementations of priority queues used
linked lists which results in $O(N)$ performance.  Implementations with
$O(N^{0.5})$ performance were introduced and analyzed in
Refs.~\cite{Blackstone, Hendriksen77,Hendriksen83,Kingston}.  The oldest
priority queue implementations with $O(\log N)$ performance used {\it
implicit heaps} binary trees in which each item always has a priority
higher than its children and the tree is embedded in an array
\cite{Bentley, Knuth, Williams}. Other $O(\log N)$ implementations
include {\it leftist trees} \cite{Knuth}, {\it binomial queues}
\cite{Brown78, Vuillemin}, {\it pagodas} \cite{Francon,Nix}, {\it skew
heaps} \cite{Sleator83,Tarjan}, {\it splay trees}
\cite{Sleator83,Sleator86,Tarjan} and {\it pairing heaps}
\cite{Fredman}.

Marin et al. \cite{Marin93,Marin} introduced a version of
the {\it complete binary tree} which also has $O(\log N)$ performance
and compared it to earlier priority queue implementations explicitly in
the context of molecular dynamics simulations.  They find that over a
wide range of densities their complete binary tree variant has the best
performance in terms of the coefficient of the $\log N$ term and large $N$
behavior.

In this work, we propose a priority queue for use in event driven 
molecular dynamics simulations for which all operations require O(1)
time.  The approach is inspired by the concept of a {\it bounded priority
queue} which is typically implemented as an array of linear lists and
which is applicable to problems in which the values associated with
queue items are integers and are bounded (i.e. the values $t$ associated
with events obey $a < t< b$ where $a$ and $b$ are constants). Bounded
priority queues are not directly applicable to the molecular dynamics
queueing problem because neither of these requirements are met.

We show, however, that with a hybrid approach that employs both a normal
priority queue and a bounded priority queue we can ensure all operations
on the queue take O(1) time.  We make use of the facts that for
molecular dynamics simulations:

(i) The time associated with an event to be added to the queue is
always later than the time associated with the last event removed from
the top of the queue.  That is,
\begin{equation}
t -t_{\rm last} \ge 0,
\label{tge}
\end{equation}
where $t$ is the time associated with the  event to be added to the queue
and $t_{\rm last}$ is the last event removed from the top of the queue.

(ii) There exists a constant, $\Delta t_{\rm max}$ such that
\begin{equation}
t-t_{\rm last}  <  \Delta t_{\rm max}.
\label{tmax}
\end{equation}
We call a priority queue which supports such events a BIPQ (Bounded
Increasing Priority Queue).

\section{Approach}

The basic idea is to:

(i) perform a gross sort of the events using an array
of linear lists and 

(ii) to use a binary tree to perform a fine sort of only those events which are
currently candidates for the event with the shortest time.

More specifically, our priority queue is composed of the following
components:

1.  An array, $A$, of $n$ linear lists $l_i$, $0 \le i < n $. (Section
\ref{choose} below discusses how to determine the the size $n$ of the
array.)  The array is treated in a circular manner.  That is, the last
linear list in the array is followed logically by the first linear list.
We implement each linear list as a doubly linked list.

2.  A binary tree which is used to implement a conventional priority queue.

We also maintain two additional quantities: the {\it current index},
$i^*$, and $i_0$ a {\it base index} associated with the queue.
Initially, all linear lists and the binary tree are empty and $i^*$ and
$i_0$ are $0$.

\section{Queue Operation}
\label{QueueOperation}

Here we describe how operations on the queue are implemented using the
data structures described above.

(i) Addition. Events are added to either one of the linear lists or to
	the binary tree as follows: An index $i$ for the event to be
	added is determined by
\begin{equation}
i = \lfloor s*t-i_0 \rfloor,
\end{equation}
where: $t$ is the time associated with the event; $i_0$ is the base
index; and $s$ is a {\it scale factor} the value of which is such the
binary tree never contains more than a relatively small number of events
( $\approx 10-20$).  If $i$ is equal to the {\it current index }, $i^*$,
the event is added to the binary tree, otherwise it is added to linear
list $l_i$.

(ii) Identification of the event with shortest time.  The event with the
shortest time is simply the root of the binary tree, as is the case with
a normal priority queue implemented using a binary tree.  If a request
is made for the event with the smallest time and the binary tree is
empty, the current index is incremented by one (wrapping around to
$i^*=0$ if we reach the end of the array) and all events in the linear
list $l_{i^*}$are inserted in the binary tree.  If there are none, we
continue to increment $i^*$ until a non-empty linear list is found.  If
we wrap around to the beginning of the list, $i_0$ is incremented by
$n$.  We find that in practice, when the binary tree becomes empty the
next linear list is always non-empty (see Section \ref{choose} in which
we show the distribution of event times).

(iii) Deletion of an event.  We simply delete the event from the array
of linear lists or from the binary tree depending on the structure in which
it is located.

The fact that the time associated with an event to be added to the queue
 is always greater than or equal to the time associated with the last event
 removed allows us to use the array of linear lists in a circular
 fashion.

The requirement that there exists a constant, $\Delta t_{max}$ such that
\begin{equation}
t - t_{\rm last} > \Delta t_{\rm max}
\end{equation}
allows us to use a finite number of linear lists. The number of linear
lists required is proportional to $\Delta t_{\rm max}$.  In practice we
find that we can always find a reasonable value of $\Delta t_{\rm max}$
such that Eq.~(\ref{tmax})~ holds.  If a rare event occurs which
violates this constraint or we want to use less memory for linear lists
causing the constraint to be violated, the event is handled on an
exception basis as implemented in the {\it processOverflowList} function
in code contained in the Appendix.  Alternatively, the application which
calls the priority queue code can guarantee that such an event never
occurs by creating an earlier fictitious collision with a time which
does not violate the constraint.

Thus all of the events, except for those deleted before they are placed
in the binary tree, will eventually be added to the binary tree, but at
any given time the tree, instead of containing O($N$) entries, will
contain only a relatively small number of entries.  The number of events
maintained in the binary tree is only a fraction of the total number of
particles $N$ in the system and can be made independent of $N$.

Our priority queue is similar to a {\it calendar queue} \cite{Brown};
 however, the calender queue does not employ a binary tree -- events are
 sorted in each of the linear lists.

\section{How to choose parameters}
\label{choose}

Two parameters, $n$ the number of linear lists and $s$ the scale factor,
must be chosen to specify the implementation of the queue.
Operationally, they can be chosen as follows:

(i) First, by instrumenting the queue to count the number of events in
the binary tree, determine a value of $s$ such that the number of events
in the binary tree is relatively small ($\approx 10-20$).  Table I. and
Fig.~\ref{mem}(a) summarize the values of $s$ we have used for our
simulations. The figure is consistent with a scale factor linear in $N$
with a different coefficient of linearity dependent on density.  Because
we use a binary tree to store events with the soonest times, the
performance of the algorithm is somewhat insensitive to the choice of
$s$.  For example, a choice of $s$ which results in a doubling of the
number of events in the binary tree results in only one additional level
in the tree.

(ii) Instrument the queue to find $\Delta t_{\rm max}$, the maximum
difference between the time associated with an event to be added and the
time associated with the last event removed and set
\begin{equation}
n=s* \Delta t_{\rm max} 
\end{equation}
to ensure that (Eq.~\ref{tmax}) is met.  Table I. and Fig.~\ref{mem}(b)
summarize the number of linear lists $n$ we have used for our
simulations. As with $s$, $n$ is linear in $N$ with a different coefficient of
linearity dependent on density.  We note that while memory requirements
are $O(N)$ as in the conventional implementation of priority queues, the
hybrid implementation does require significantly more memory than the
conventional implementation due to the memory required for the linear
lists.  Tradefoffs can be made of cpu time for memory by increasing the
scale factor and/or reducing the number of linear lists (resulting in
more exception conditions).

Figure {\ref{pdist}(a) plots $\langle m_{\hat i} \rangle$ the average
number of events with index $\hat i$ versus $\hat i$ for various $N$.
Here
\begin{equation}
\hat i\equiv (i-i^*+n) \mod n.
\end{equation}
That is, $\hat i$ is the distance of $i$ from the current index taking
into account the circular nature of the array of linear lists.  The data
was obtained by sampling the queue many times at regular intervals.
With the choice of scale factors shown in Table I we achieve our goal
of having $\approx 10-20$ events with index $i=i^*$ and thus in the
binary tree.  Note that to achieve this, the scale factor increases with
increasing $N$ resulting in the cutoff of the distributions also
increasing with increasing $N$.  (In fact, if the $x$-axis is
transformed by $x=x/N$, the plots collapse as shown in
Fig.~\ref{pdist}(b) reflecting the fact that the probability
distribution of collision times is independent of $N$.)  Thus, the
number of linear lists required to ensure that Eq.~(\ref{tmax}) holds
also increases with $N$.  In Fig.~\ref{pBucket0}, we plot the
distribution $P(m^*)$ the probability that the number of events with
the current index, $i^*$ is $m^*$ versus $m^*$.  The distributions are
strongly peaked indicating that the number of events in the binary tree
do not vary much from the average.

\section{Complexity Analysis}
\label{complexity}

The basic operations involved in the queue are:

(i) insertion into and deletion from the linear lists.  Use of doubly
linked lists allows these operations to be implemented to take O(1) time.

(ii) binary tree operations.  We use the code of Ref.~\cite{Marin} to
implement the binary tree operations.  When a leaf representing an item
in the priority queue is added to or deleted from the tree, the tree
must be traversed from the affected leaf possibly all the way to the
root node and adjustments made to reflect the presence or absence of the
affected leaf.  Thus a bound on the number of levels which must be
traversed is $\log_2 m$ where $m$ is the number of items in the priority
queue.  In Sec.~ \ref{choose} we show that by choosing the scale
factor $s$ appropriately, $m$ can be made to be independent of $N$ (and
have a relatively small value, $\approx 10-20$).  Thus binary tree
operations will be O(1).

(iii) identification of the next non-empty linear list, after the
current linear list is exhausted.  As explained in item (ii) of
Sec. \ref{QueueOperation}, when the binary tree is empty, we search
forward through the array of linear lists until a non-empty list is
found.  If the number of lists we must search through increases with
$N$, this process will not be O(1).  We show below that with the proper
choice of $s$, the number of lists we must search does not grow with $N$
and in fact show that the next linear list after the current one almost
always is non-empty.  Thus the complexity of identification of the next
non-empty list will be O(1).

Thus the overall time taken by queue operations per collision is O(1).

\section{Experiments /Simulations }

We run simulations using both a conventional priority queue and our new
 hybrid approach.  For simplicity the simulation was of identical size
 hard spheres of radius one and unit mass.  The sizes $L$ of the cubic
 systems are set to maintain equal densities.  The parameters of the
 simulation are as shown in Table I.

To demonstrate the performance of our approach, we run simulations for
cubic systems at four volume densities $\rho=0.01, 0.12, 0.40$ and $0.70$.
The first density represents a rarefied gas and the last density
represents a jammed system.  The jamming density for hard sphere systems
is $\approx 0.64$ \cite{Donev2004}. For both the conventional priority
queue and the hybrid queue we used the binary tree code from
Ref.~\cite{Marin}.

Figure \ref{pcbt} shows the time taken for  $10^7$ collisions for queue
operations with both a conventional priority queue and the hybrid queue.
As expected, the time for the conventional priority queue increases as
$\log N$ while the time for the hybrid queue is essentially constant.

There is, however, a slight upward trend in the hybrid queue results.
To determine if this trend is a feature of the algorithm or of the
the benchmark environment, we proceed as follows.

We first study the only two places in the hybrid code where looping is
involved:

(i) in the {\it updateCBT} function of Ref. ~\cite{Marin} we loop as we
traverse the binary tree.  If we traverse more levels as the N grows, the
algorithm will not be O(1).  To explore this possibility, we instrument
the function to count the number of number of levels we traverse in the
tree.  The results are shown in Fig.~ \ref{pcbt}.  The number of loop
iterations is essentially constant, independent of $N$.

(ii) in the {\it deleteFirstFromEventQ} function, after the
priority queue for the current linear list is exhausted, we loop until we
find a non-empty list.  If the number of lists we must examine before we
find the first non-empty list grows with the system size, the algorithm
will not be O(1).  We examine this possibility by counting the number of
times we encounter an empty list and find that on average the
probability of encountering an empty list does not grow with $N$ and
that the probability of encountering an empty list is very small: we
encounter an empty list only $10^{-4}$ of the times after exhausting the
priority queue.

Having ruled out dependence of the number of loop iterations on $N$ as
the source of the upward trend in the execution times, we now consider
whether the larger memory needed as $N$ increases is the cause of the
trend.  All modern computer processors employ high speed cpu cache
memory to reduce the average time to access memory
\cite{Handy,Hennessy}.  In fact, the processor we use in our
simulations, the AMD Opteron, employs a two-level memory cache (64 KB
level 1 cache, 1 MB level 2 cache) \cite{Opteron}.  A similar cache
structure is used in the Intel Xeon processor \cite{Xeon}.  Because
memory caches are finite size, if the memory access is random the larger
the memory used by a program, the lower the probability that data will
be found in the cache resulting in slower instruction execution.  The
effect of cache in benchmark runtimes has been studied in
Ref. ~\cite{Saavedra}.  We study the effect of the finite size of the
cache in our system as follows: Instead of running the molecular
dynamics simulations, we run a small test program which randomly
accesses the data structures used by the molecular dynamics simulations.
For each value of $N$, the test program executes exactly the same number
of instructions but uses data structures of the size used by the
molecular dynamics simulations for that value of $N$.  The results are
shown in Fig.~\ref{pcbt} and show an upward trend similar to that of the
simulation results for all of the densities studied.

The above results thus suggest that the complexity of the hybrid
algorithm is, in fact, O(1) and that the upward trend in the results is
due to the finite size of the high speed memory cache.

\section{Discussion and Summary}

We have defined a new abstract data type, the Bounded Increasing
Priority Queue (BIPQ) having the same operations as a conventional
priority queue but which takes advantage of the fact that the value
associated with an item to be added to the queue has the properties that:
(i) the value is greater than or equal to the value associated with the
last item removed from the top of the queue and (ii) the value minus the
value of the last item removed from the top of the queue is bounded.
These properties are obeyed for events in event driven molecular dynamic
simulations.  We implement a BIPQ using a hybrid approach incorporating
a conventional priority queue (implemented with a binary tree) and a
bounded priority queue.  All operations on the BIPQ take an average O(1)
time per collision. This type of queue should provide performance
speedups for molecular dynamics simulations in which the event queue is
the bottleneck.

\section{Acknowledgments}

We thank Sergey Buldyrev, Pradeep Kumar, Sameet Sreenivasan, and Brigita
Urbanc for helpful discussions.  We ONR, NSF and NIH for support.

\newpage
\appendix
\begin{center}
\bf {APPENDIX}
\end{center}

The following code implements the hybrid queue proposed here.  The calls
to Insert and Delete are to the functions contained in
Ref.~\cite{Marin}, which update NP and the complete binary tree, CBT.
Any code providing the same functions could be substituted for Insert
and Delete.

\begin{singlespace}
\begin{verbatim}

#define nlists 50000
#define scale 50

typedef struct
{
  int next;			
  int previous;	
  int p;
  int q;
  int c;
  double t;
  unsigned int qMarker;
  int qIndex;
  statusType status;
}eventQEntry;

eventQEntry * eventQEntries;
double baseIndex;

int * CBT;  /* complete binary tree 
              implemented in an array of 
              2*N integers */
int NP;     /*current number of particles*/

int linearLists[nlists+1];/*+1 for overflow*/
int currentIndex;


//----------------------------------------
int insertInEventQ(int p)
{
  int i,oldFirst;
  eventQEntry * pt;
  pt=eventQEntries+p;   /* use pth entry */

  i=(int)(scale*pt->t-baseIndex);
  if(i>(nlists-1))     /* account for wrap */			
  {
    i-=nlists;
    if(i>=currentIndex-1)
    {      
      i=nlists;	/* store in overflow list */
    }
  }
  pt->qIndex=i;

  if(i==currentIndex)
  {
    Insert(p);   /* insert in PQ  */
  }
  else
  {
    /* insert in linked list */

    oldFirst=linearLists[i];
    pt->previous=-1;	
    pt->next=oldFirst;					
    linearLists[i]=p;

    if(oldFirst!=-1)
      eventQEntries[oldFirst].previous=p;	
   }
    return p;
}

//----------------------------------------

processOverflowList()
{
  int i,e,eNext;
  i=nlists;  /* overflow list */
  e=linearLists[i];
  linearLists[i]=-1;  /* mark empty; we will
     treat all entries and may re-add some */
  while(e!=-1)
  {
    eNext=eventQEntries[e].next; /* save next */
    insertInEventQ(e);	/* try add to regular list now */
    e=eNext;
  }
}

//---------------------------------------

void deleteFromEventQ(int e)
{
  int prev,next,i;
  eventQEntry * pt=eventQEntries+e;

  i=pt->qIndex;
  if(i==currentIndex)	
    Delete(e);   /* delete from pq */
  else				
  {
    /* remove from linked list */

    prev=pt->previous;
    next=pt->next;	
    if(prev==-1)
      linearLists[i]=pt->next;
    else
      eventQEntries[prev].next=next;

    if(next!=-1)
      eventQEntries[next].previous=prev;
  }
}

//---------------------------------------

int deleteFirstFromEventQ()
{
  int e;

  while(NP==0)/*if priority queue exhausted*/
  {
    /* change current index */

    currentIndex++;
    if(currentIndex==nlists)
    {
      currentIndex=0;
      baseIndex+=nlists;
      processOverflowList();
    }

    /* populate pq */

    e=linearLists[currentIndex];
    while(e!=-1)
    {
      Insert(e);
      e=eventQEntries[e].next;
    }
    linearLists[currentIndex]=-1;
  }

  e=CBT[1];    /* root contains shortest
                  time entry */
	
  Delete(CBT[1]);
  return e;
}
//---------------------------------------


\end{verbatim}
\end{singlespace}

\begin{table}[htb]
\caption{Parameters of molecular dynamics simulations.}
~
\begin{tabular}{ccc}

 {Number} &  {Scale Factor}& {Number} \\
 {of Particles}  & & {of lists} \\
~
 $N$      & $s$ & $n$ \\  \hline
\multicolumn 3 c {$\rho=0.01$} \\  \hline
$1000$ &  $100$ &  $25000$ \\
$8000$ &  $700$ &  $200000$    \\
$64000$ & $5000$  &   $2.5 \times 10^6$ \\ 
$512000$ &  $45000$ &   $25 \times 10^6$ \\  \hline
\multicolumn 3 c {$\rho=0.12$} \\  \hline
$1000$ &  $50$ &  $50000$ \\
$8000$ &  $500$ &  $400000$    \\
$64000$ & $3400$  &   $5 \times 10^6$ \\ 
$512000$ &  $25000$ &   $50 \times 10^6$ \\  \hline
\multicolumn 3 c {$\rho=0.4$} \\  \hline
$1372$ &  $1000$ &  $250000$ \\
$8788$ &  $7500$ &  $2 \time 10^6$    \\
$70304$ & $60000$  &   $16 \times 10^6$ \\ 
$530604$ &  $500000$ &   $130 \times 10^6$ \\  \hline
\multicolumn 3 c {$\rho=0.7$} \\  \hline
$1372$ &  $15000$ &  $500000$ \\
$8788$ &  $75000$ &  $200000$    \\
$70304$ & $500000$  &   $35 \times 10^6$ \\ 
$530604$ &  $4 \times 10^6$ &   $300 \times 10^6$ \\  \hline
\end{tabular}
\end{table}


\begin{figure}
\centerline{
\xsize
\epsfclipon
\epsfbox{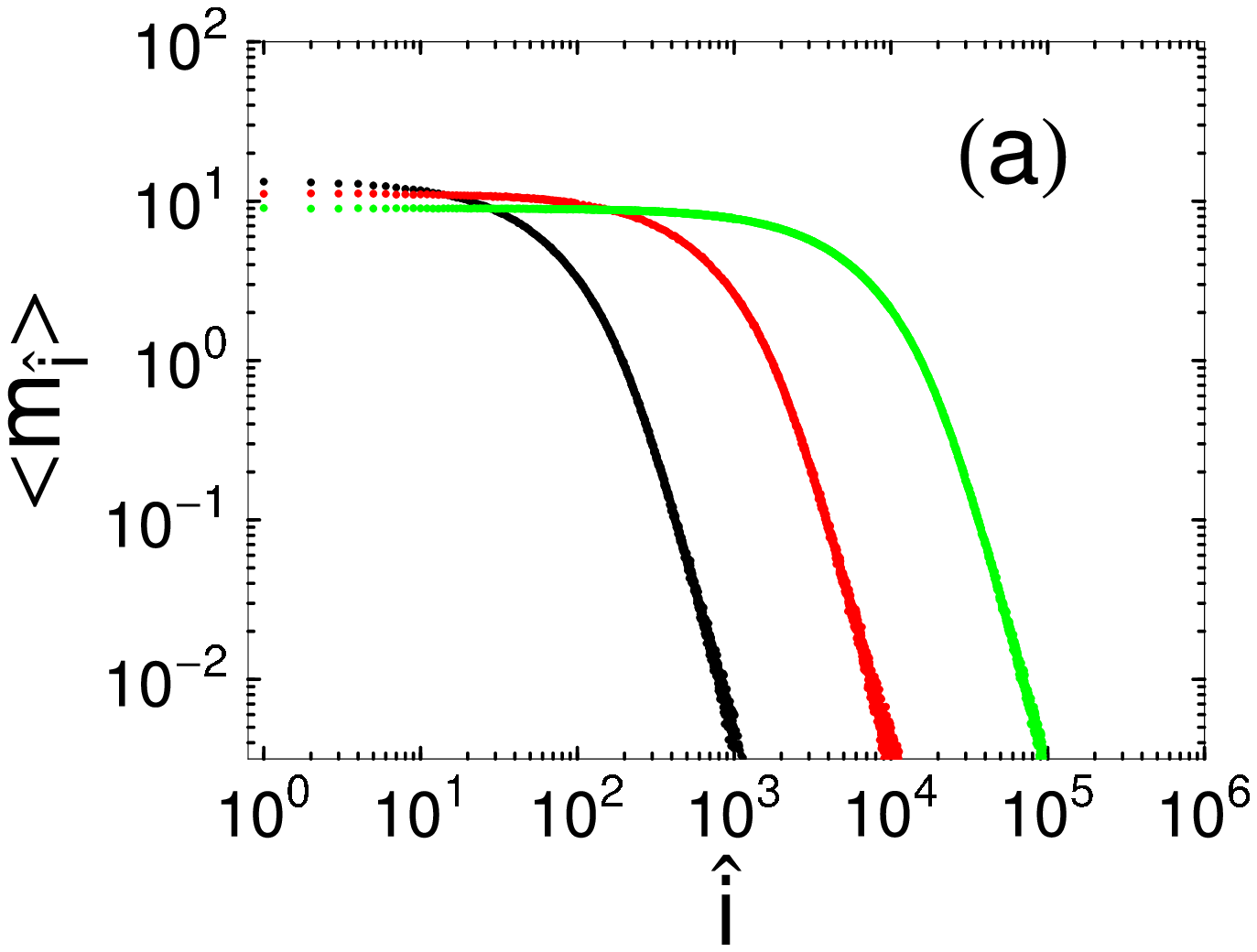}
}

\centerline{
\xsize
\epsfclipon
\epsfbox{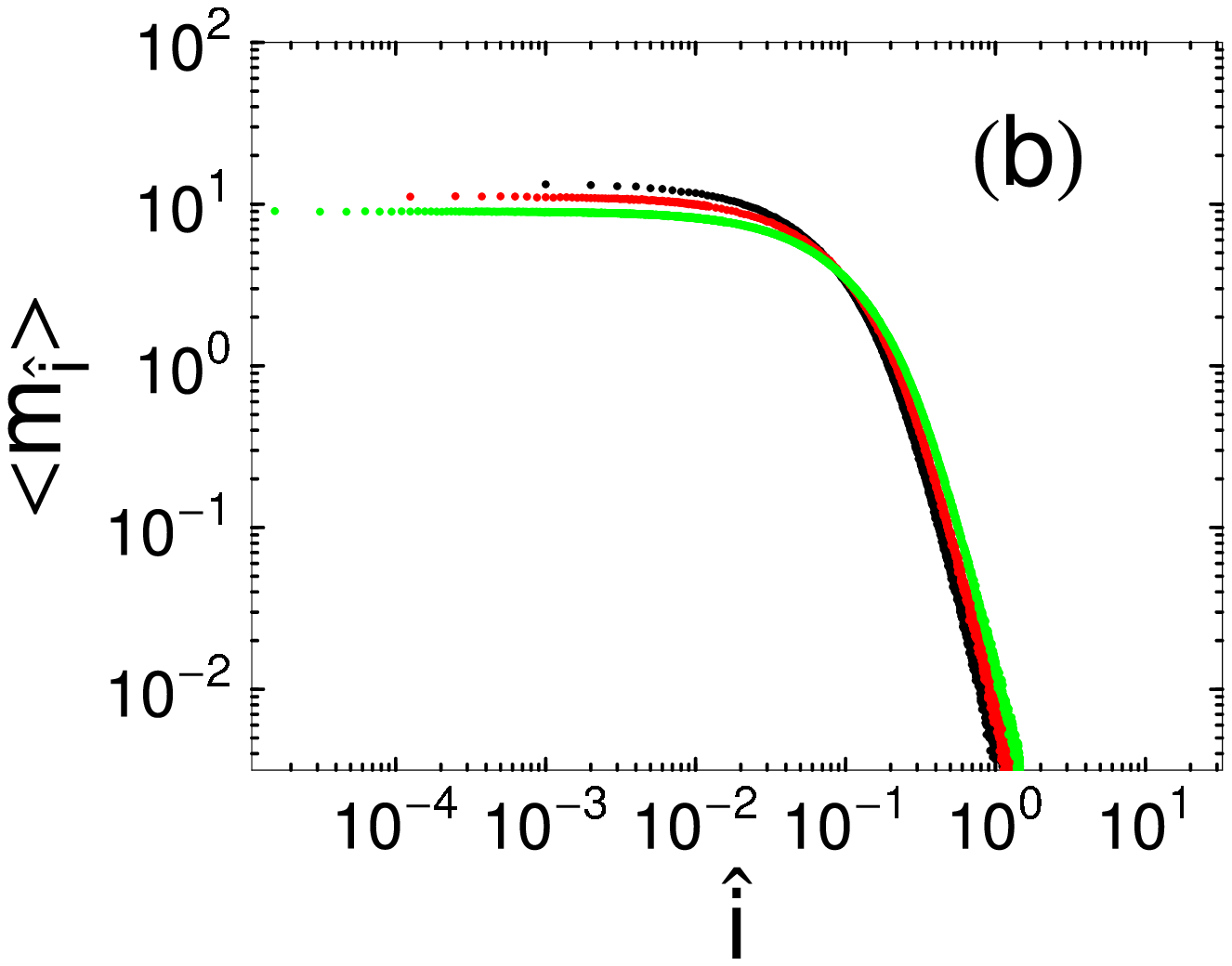}
}

\caption{(a) For $\rho=0.12$, the average number of events $\langle m_{\hat i} \rangle$
  with index $\hat i$ versus $\hat i$(the distance of $i$ from the
  current index $i^*$) for (from left to right $N$=1000, 8000, and 64000.
  (b) Same as (a) with the x-axis scaled by $1/N$ which results in a
  collapse of the plots. }
\label{pdist}
\end{figure}

\begin{figure}
\centerline{
\xsize
\epsfclipon
\epsfbox{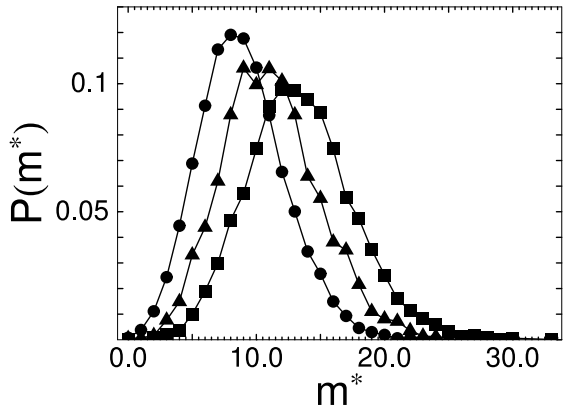}
}
\caption{For $\rho=0.12$, $P(m^*)$, the probability that the number of events in linear
list $i=0$ is $m^*$, vs. $m^*$ for $N=$1000(squares), 8000(triangles),
and 64000(disks).  }
\label{pBucket0}
\end{figure}

\begin{figure}
\centerline{
\xsize
\epsfclipon
\epsfbox{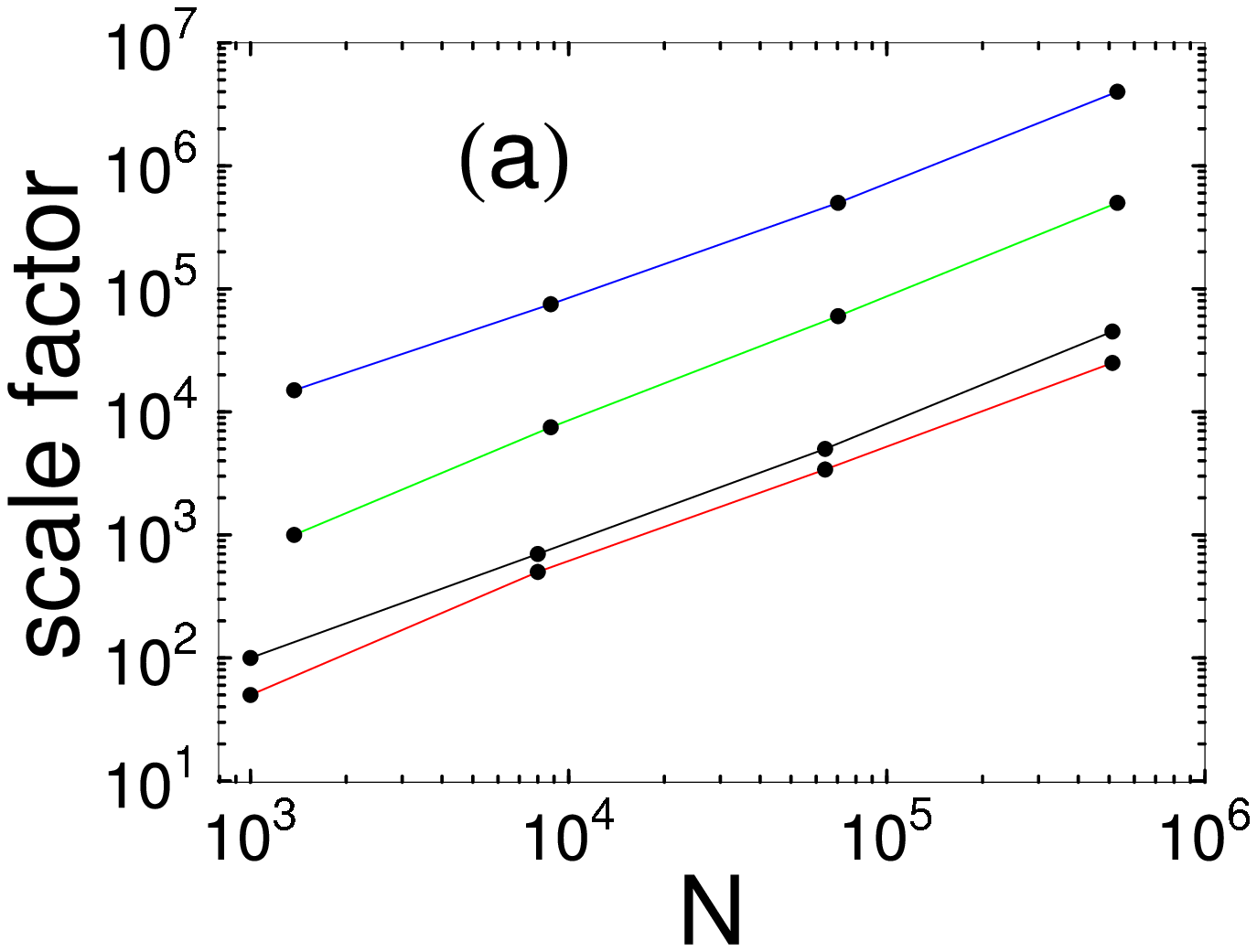}
}
\centerline{
\xsize
\epsfclipon
\epsfbox{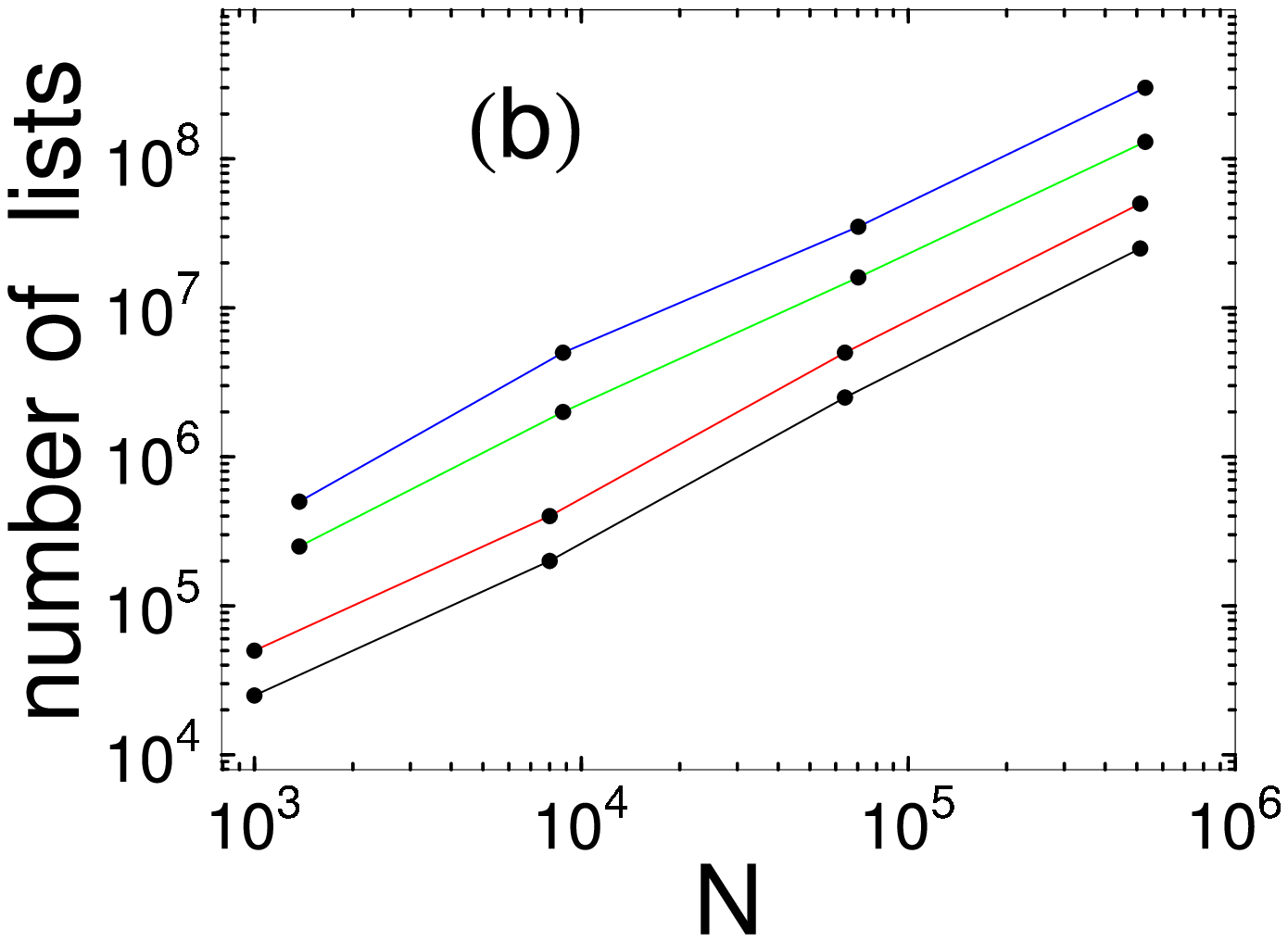}
}

\caption{(a) Scale factor, $s$, vs $N$ for (from bottom to top)  $\rho=0.12,
 0.01, 0.4$ and $0.7$. (b) Number of linear lists, $n$, vs $N$ for (from bottom to top)  $\rho=0.01,
 0.12, 0.4$ and $0.7$.} 
\label{mem}
\end{figure}

\begin{widetext}

\begin{figure}
\centerline{
\xsize
\epsfclipon
\epsfbox{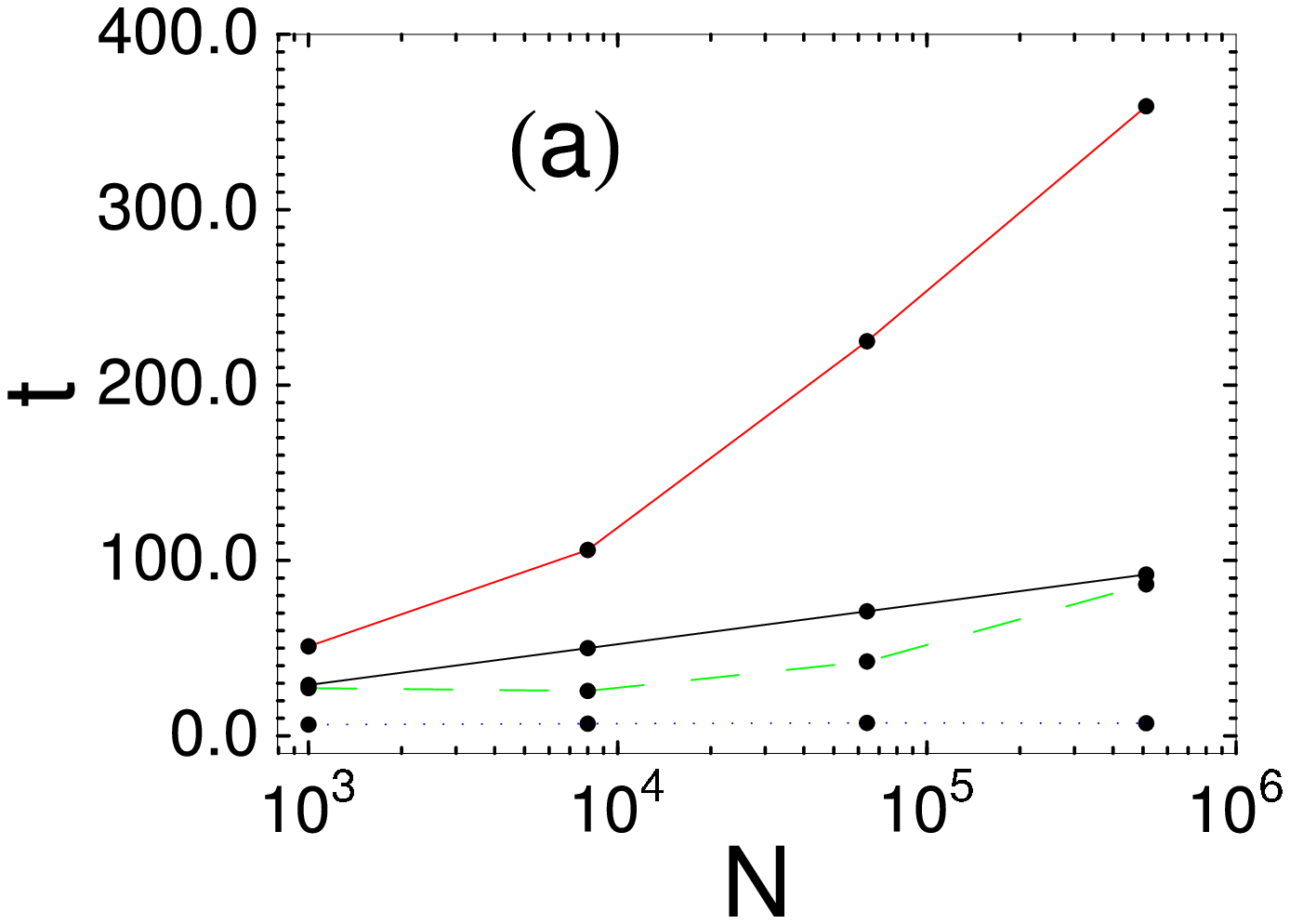}
\xsize
\epsfclipon
\epsfbox{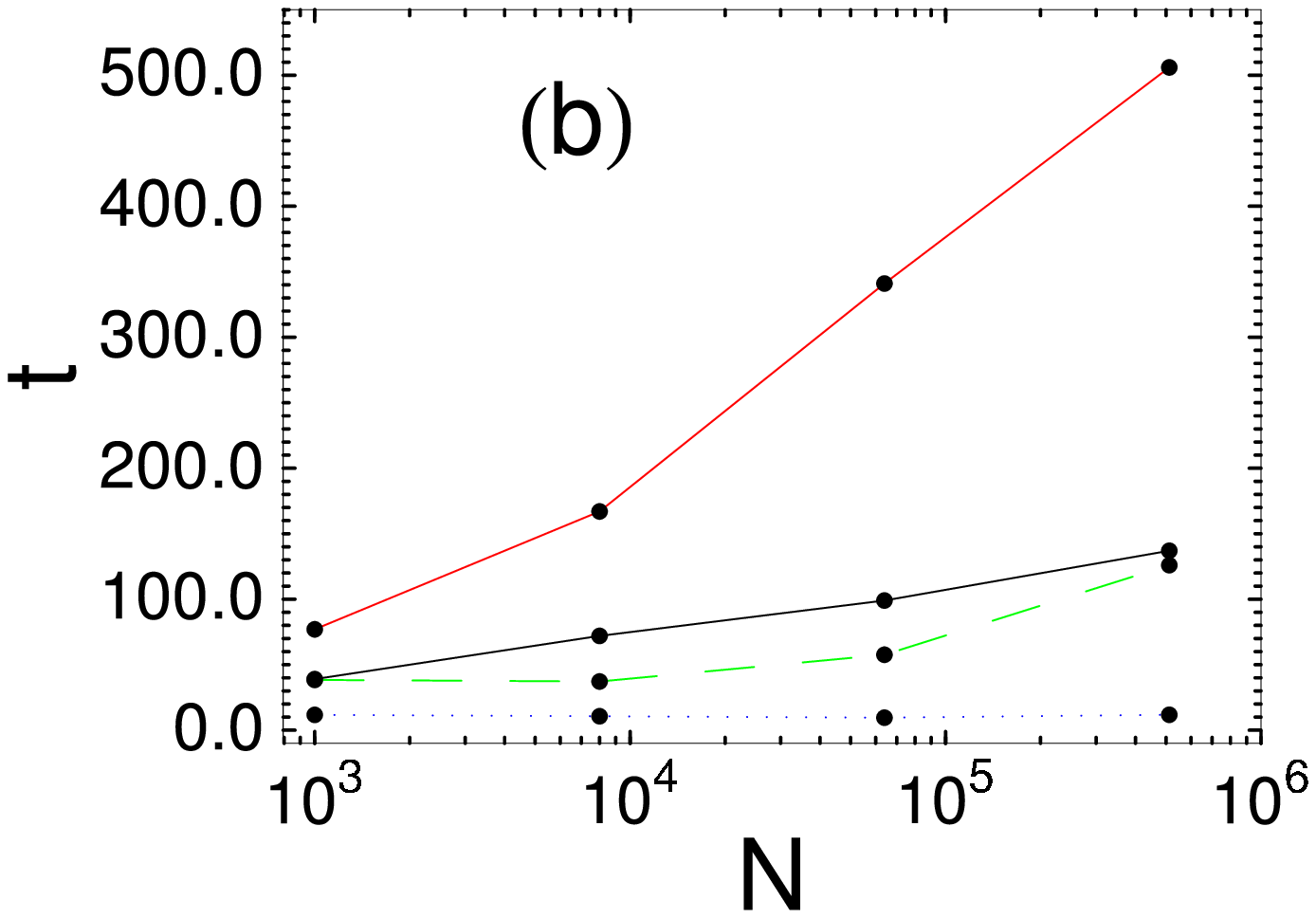}
}

\centerline{
\xsize
\epsfclipon
\epsfbox{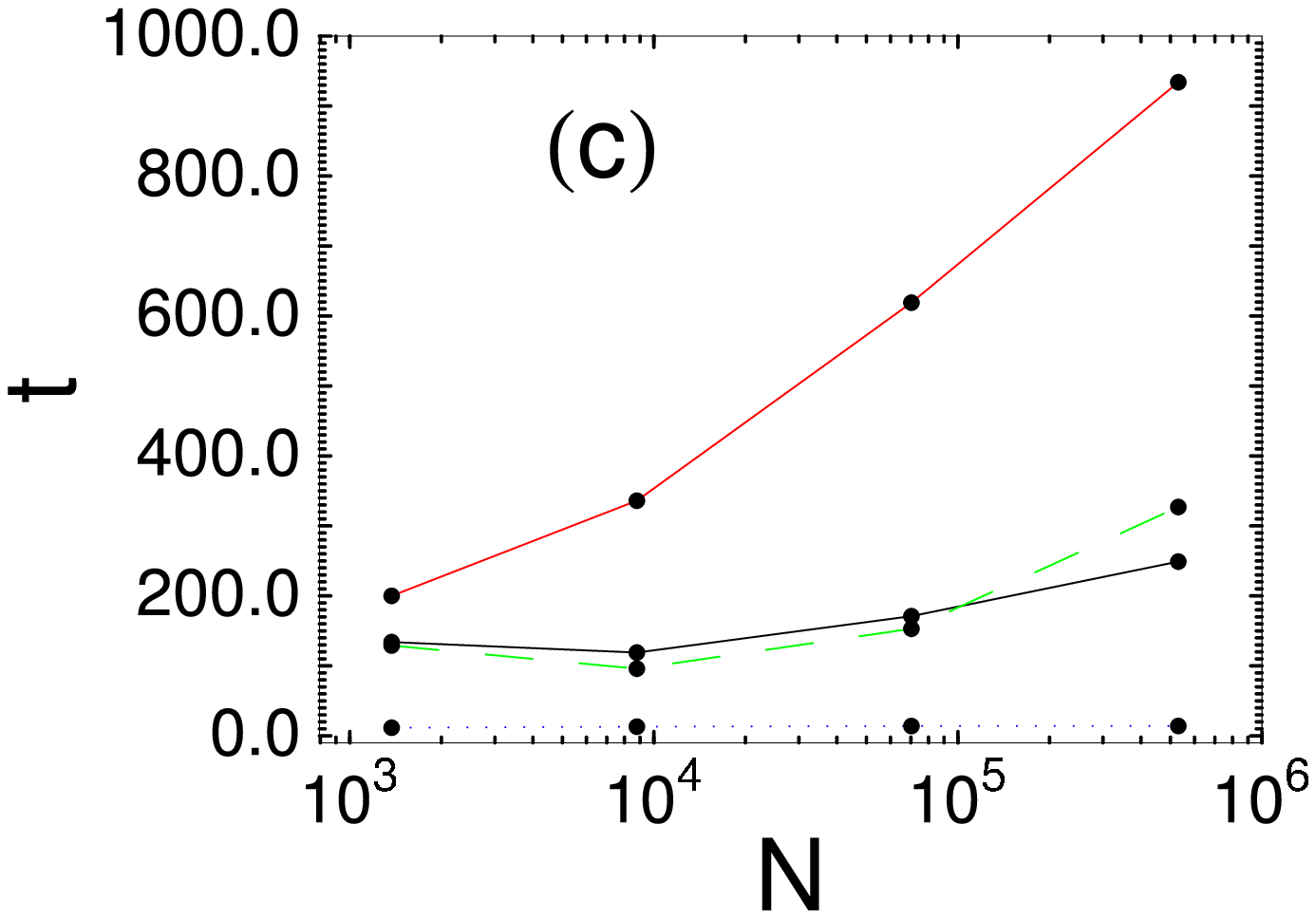}
\xsize
\epsfclipon
\epsfbox{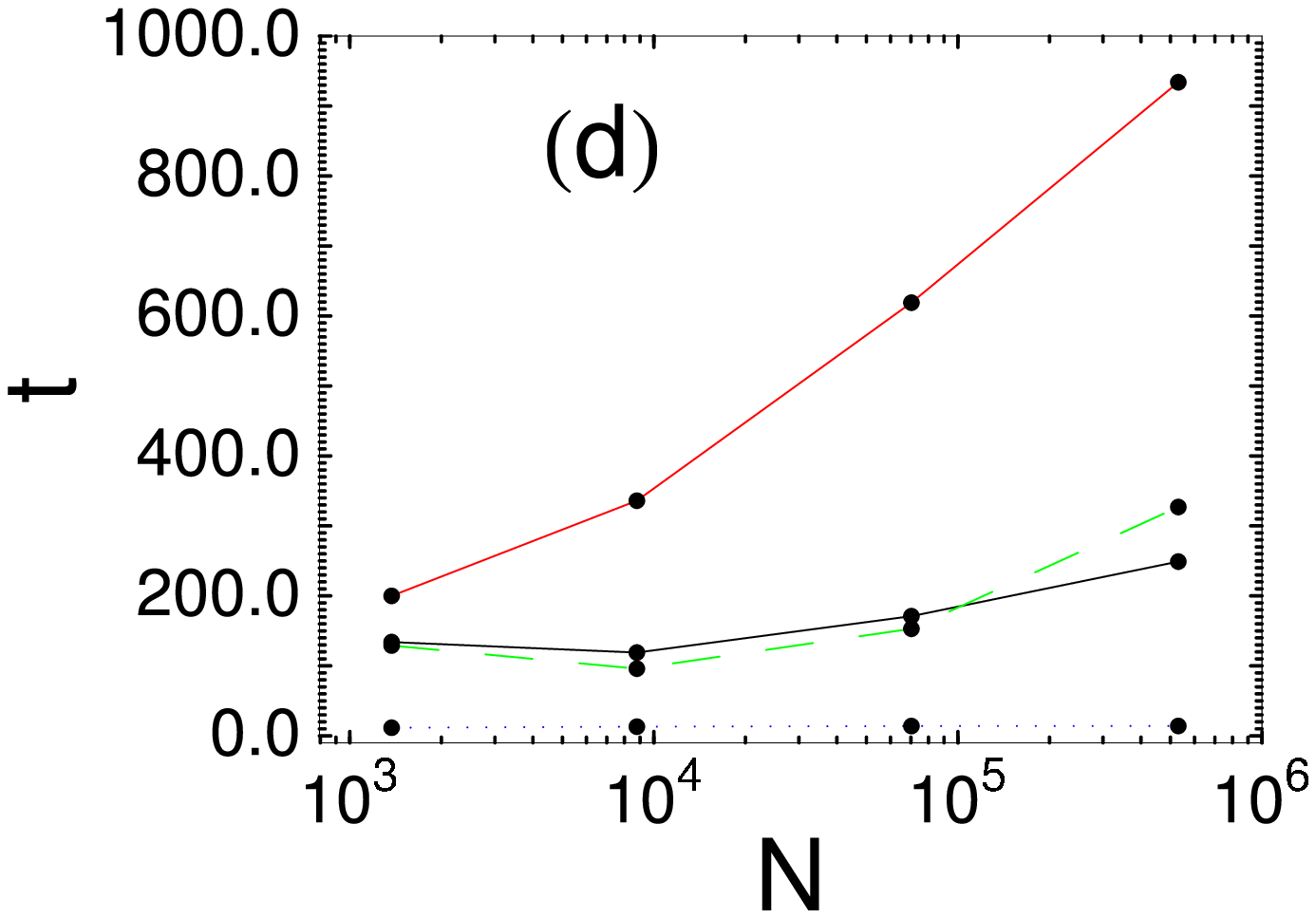}
}

\caption{Processing time for queue operations vs. $N$, the number of
particles in the system.  (a) Volume density $ \rho =0.01$.  The higher
solid line is the processing time for queue operation for a normal
priority queue; the lower solid line is for the hybrid queuing system
introduced here.  The dashed line represents the benchmark test timing
to execute a fixed number of instructions independent of $N$ but with
memory sizes corresponding to the memory used for the hybrid system.
The dotted line represents the number of tree levels traversed ($ \times
10^{-7})$ in the binary tree for the hybrid system.  (b),(c) and (d)
Same as (a) for $\rho=0.12, 0.4$, and $0.7$, respectively.}
\label{pcbt}
\end{figure}

\end{widetext}
}


\end{document}